\documentclass[preprint,amsmath,amssymb]{article}

\usepackage{dcolumn}
\usepackage{bm}
\usepackage{hyperref}
\usepackage{amsmath}
\usepackage{amsfonts}
\usepackage{graphics}
\usepackage{graphicx}
\usepackage{epsfig}
\usepackage{rotating}
 \usepackage[latin1]{inputenc}
 \usepackage{color}
 \usepackage{rotating}
 \usepackage{amssymb}
\usepackage{epsfig}
\usepackage{textcomp}
 \usepackage{setspace}
\usepackage{natbib}

\begin{document}

\begin{center}
{\Large \bf
Fixation of a Deleterious Allele under  Mutation Pressure and Finite Selection Intensity}
\end{center}

\vspace*{0.5cm}

\begin{center}
{\large Michael Assaf$^{+,*}$ and Mauro Mobilia$^{\dagger,*}$}

\vspace*{0.4cm}

{\small
$^{+}$ University of Illinois at Urbana--Champaign\\
Loomis Laboratory of Physics, Department of Physics\\
1110 West Green Street, Urbana, Illinois 61801, USA\\
Phone:~+1-217-333-0929\\
Email:~assaf@illinois.edu
}

\vspace*{0.4cm}

{\small
$^{\dagger}$ University of Leeds\\
Department of Applied Mathematics, School of Mathematics\\
Leeds LS2 9JT, United Kingdom\\
Phone:~+44-(0)11-3343-1591\\
Email:~M.Mobilia@leeds.ac.uk
}

\vspace*{0.4cm}

{\small
$^{*}$ Both authors contributed equally to this work.
}

\end{center}

\newpage

{\bf Abstract}\\
The mean fixation time of a deleterious mutant allele
is studied beyond the diffusion approximation.
As in Kimura's classical work [M. Kimura, Proc. Natl. Acad. Sci. U.S.A. {\bf 77},
522 (1980)], that was motivated by the problem of fixation in the presence of
amorphic or hypermorphic mutations,  we consider a diallelic model at a single locus comprising a wild-type $\textsf{A}$
and a mutant allele $\textsf{A'}$ produced irreversibly from $\textsf{A}$ at small uniform rate $v$.
The relative fitnesses of the mutant homozygotes $\textsf{A'} \textsf{A'}$,
mutant heterozygotes $\textsf{A'} \textsf{A}$ and wild-type  homozygotes $\textsf{A} \textsf{A}$
are $1-s$, $1-h$ and $1$, respectively, where it is assumed that $v\ll s$.
Here, we employ a WKB theory and directly treat the underlying Markov chain (formulated as a birth-death process)
obeyed by the allele frequency (whose  dynamics is prescribed by the Moran model). Importantly, this approach allows to accurately account for
 effects of large fluctuations.
After a general description of the theory, we focus on the case of a deleterious
mutant allele (i.e. $s>0$) and discuss three situations: when the mutant is (i) completely dominant ($s=h$);
 (ii) completely recessive ($h=0$), and (iii)  semi-dominant ($h=s/2$).
Our theoretical predictions for the mean fixation time and the
quasi-stationary distribution of the mutant population in the coexistence state,
are shown to be in excellent agreement with numerical simulations. Furthermore, when $s$ is finite, we demonstrate that our results are
superior to those of the diffusion theory, while the latter is shown to be an accurate approximation
only when $N_e s^2\ll 1$, where $N_e$ is the effective population  size.
\\
\vspace*{0.3cm}


\newpage

\section{Introduction}
 Fixation is a phenomenon that allows to quantify how the frequency dynamics of a rare allele increases under
genetic drift and selective forces until it takes over the entire population, and is a key topic in population genetics. In particular, the fixation of alleles is crucial
to understand genetic diversity and
has therefore
 attracted much attention since the pioneering works by Fisher and Wright (\cite{Fisher,PopGen,Wright31}).
The study of the fixation time of a deleterious allele
has been shown to be important to shed light on the evolutionary dynamics of related diseases~(\cite{Slatkin}).
In addition to selection and genetic drift, mutation is another important mechanism responsible of genetic diversity.
In each generation, mutant genes arise naturally and can be either deleterious, selectively neutral or advantageous.
A vast majority of mutants are lost in a few generations and only a small number of them succeed in fixating the population~(\cite{KimuraOhta}).
As already noticed by Fisher~(\cite{Fisher}) a majority of mutations with large effect are deleterious and in these cases it is almost certain that the mutant allele will be eliminated before reaching an appreciable frequency. However, it has been reported that for some amorphic or hypomorphic mutations the
previously deleterious mutations may become only slightly deleterious (i.e. almost neutral)
and, under mutation pressure and random drift,  become fixed in
the species~(\cite{Muller1939,Wright1977,Kimura1980}). Examples of amorphic/hypomorphic  deleterious mutations
range from the loss of eyes in cave
animals to the loss of ability to synthesize vitamin C in some vertebrates~(\cite{Jukes1975}).

In this context, understanding the combined influence of genetic drift, mutations and natural selection
on the dynamics of the allele frequency  is
clearly an issue of fundamental importance in evolutionary biology~(\cite{CrowKimura,KimuraOhta,Kimura,Ewens,Gillespie04}).
A large body of studies dedicated to gene frequency is based
on the diffusion approximation [see, e.g.,~(\cite{CrowKimura,Ewens})]. The latter approach is intrinsically a neutral theory
where the allele frequencies are treated as continuous random variables. The diffusion theory
 has been extremely insightful as, for instance, it has allowed
 the computation of the time-dependent probabilities of fixation
of a gene under selection in an ideal (randomly-mating) diploid population of large effective size. The diffusion theory has been recently generalized to study
fixation in spatially structured metapopulations (\cite{Whitlock}) and has been used to statistically test the existence of
positive selection~(\cite{Zeng1,Zeng2})~\footnote{In those references, the frequency trajectories of
the derived alleles are generated using the pseudo-sampling method of Ref.~(\cite{Kimura1980}), see Appendix.}.
The diffusion approximation has also been used to study fixation in quasi-neutral systems where the intensities of selection
and mutations are much weaker than the genetic drift's strength.
In fact, if ${\widetilde s}$ denotes
the typical strength of non-neutral evolutionary forces (selection and mutation)
in a population of (effective) size ${\cal N}$,  the diffusion  approximation certainly
gives accurate results when ${\cal N}{\widetilde s}$ is a small quantity~(\cite{CrowKimura,Ewens}).
Using heuristic arguments, Nei has recently
 proposed to replace the above criteria
by a less stringent requirement ${\widetilde s}< {\cal N}^{-1/2}$~(\cite{Nei}).
The detailed analytical and numerical
analysis presented in this work substantiates that ${\cal N}{\widetilde s}^2\ll 1$
is indeed the necessary condition for the predictions
of the diffusion approximation to be adequate. While the diffusion approximation is valid for a narrow range of vanishingly small selection intensity,
it is commonly employed also when ${\cal N}{\widetilde s}^2\gtrsim 1$. As an example, it is often considered that
  ${\widetilde s}\approx 0.1\% - 1\%$  and ${\cal N}\approx 10^{4}
- 10^{5}$~(\cite{KimuraOhta,CrowKimura,Ewens}) which yields finite values of ${\cal N}{\widetilde s}^2$ ranging
from $0.01$ to $10$. In this situation, the assumption of quasi-neutrality, and therefore the validity of the diffusion approximation,
appears to be questionable. Nevertheless,
 to our knowledge there have been no systematic investigations to
establish the validity of the diffusion approximation, and to improve over it,
 in systems with arbitrary (finite) selection  intensity and
mutations~\footnote{Only recently, it has been shown that some results of
the diffusion approximation are prone to numerical inaccuracies~(\cite{Wang}).}.

In this paper, we accurately compute the mean fixation time and the quasi-stationary properties of a deleterious allele in a
 panmictic population of diploid individuals under mutation pressure and finite selection intensity.
While Kimura and other authors~(\cite{Kimura1980,Li}) considered this problem within the realm of diffusion approximation,
we here adopt a different approach originating from statistical physics. By doing so, we are able to capture {\it non-Gaussian} and {\it large-fluctuation} effects that govern the evolutionary dynamics when the selection strength
is {\it nonvanishingly weak}.
Indeed, it has been recently noticed that methods borrowed from statistical physics can be extremely insightful
to address problems of evolutionary dynamics from a broader perspective~(\cite{Sella,Korolev}). Here,
we employ a so-called WKB theory (see below) to obtain the
mean-fixation time (MFT) directly from the Markov chain (formulated as a birth-death process, see Sec.~2.2) governing the stochastic dynamics of the allele frequency.
This method, that relies on a power series expansion in the inverse of the (effective) population size and a suitable ansatz (i.e. a trial function),
allows us to accurately account for {\it non-diffusive} phenomena like
fixation that are triggered by rare large fluctuations. As  main results, in this work  (i) we
demonstrate that the diffusion approximation is adequate only when $s\ll {\cal O}({\cal N}^{-1/2})$ and
${\cal N}\gg 1$,
 corroborating Nei's heuristic estimate
(\cite{Nei}); (ii) we derive accurate results  for the MFT and the population's quasi-stationary distribution
(QSD) when $s\gtrsim {\cal O}({\cal N}^{-1/2})$.

The organization of this paper is the following: the model and the methods are presented in the next section. There, we first discuss
the deterministic dynamics obtained when fluctuations are ignored, and then derive a stochastic description in terms
of a continuous-time birth-death process (using the fitness-dependent
Moran model) that takes demographic stochasticity into
 account. Our (WKB-based) analytical approach is then presented in
Section 3.
The results for the MFT and the QSD are presented and discussed in Sections 4 and 5.1.
Section 5.2 is dedicated to a careful comparison of our results with the results of the diffusion approximation and is
followed by a discussion and our conclusions. The numerical simulation method is briefly described in an Appendix.

\section{Model and Methods}
We consider a diallelic model of randomly mating population of $N$ diploid individuals with
an effective size $N_e$~\footnote{$N_e$ can essentially be interpreted as the number of individuals that can breed in
each generation, see e.g. Refs.~(\cite{CrowKimura,Ewens,Gillespie04}).}. We assume that at a particular locus, where the wild-type allele
is $\textsf{A}$ while the mutant is $\textsf{A}'$, the relative fitnesses of the homozygotes $\textsf{A}\textsf{A}$,
$\textsf{A}'\textsf{A}'$ and heterozygotes $\textsf{A}'\textsf{A}$ are respectively  $1$, $1-s$ and $1-h$
(with $s>0$ and $h\geq 0$). This corresponds to a generic diallelic model where the
mutant allele is deleterious~(\cite{KimuraOhta,CrowKimura,Ewens,Gillespie04}).
  Following (\cite{Kimura1980,Li}), in addition to random mating
we also assume that the wild type  mutates irreversibly to the  (deleterious)
type $\textsf{A}'$ with a small rate $v$, which accounts for amorphic and hypomorphic mutations
(\cite{Kimura1980}).  In reality, the allele  $\textsf{A}$ mutates in various types that are here regarded as a single class denoted by $\textsf{A}'$.

A situation of particular relevance in population genetics arises when the selection intensity $s$ is weak and
the mutation rate $v$ is much weaker, i.e. $0<v\ll s \ll 1$, where $v$ is  several order of
magnitudes smaller than $s$~(\cite{Ewens,CrowKimura,Kimura,KimuraOhta})~\footnote{Typically,
$N_e\approx 10^4 - 10^5, \, s\approx 10^{-4} - 10^{-2}$ and $v \approx 10^{-4} - 10^{-6}$. Thus,
$N_e s \approx 1-1000$, $N_e v \approx 0.01 - 10$, whereas $N_e s^2 \approx 10^{-4} - 10$ and
$N_e v^2 \approx 10^{-8} - 10^{-3} \ll 1$.}.
The MFT of this model where the mutant allele $\textsf{A}'$ is either dominant, recessive or semi-dominant
 was treated both analytically and numerically within the diffusion approach in (\cite{Kimura1980}), while
the non-dominant case was considered in  (\cite{Li}).
Here, we are are mainly interested in the MFT and QSD
of the allele frequencies
 in the presence of
(small but) non-vanishing selection intensity and mutation pressure, when the parameters are such
that $0<v\ll s\ll 1$, $ s \gtrsim{\cal O}(N_e^{-1/2})$ and the validity of the diffusion approximation is thus questionable.
In fact, due to the broad use of the diffusion theory, an important question
concerns how the selection intensity  $s$ should vanish with the effective population size $N_e$
for the diffusion approximation to be valid, and we here demonstrate that the condition to be satisfied in fact is $s\ll {\cal O}(N_e^{-1/2})$.

\subsection{The deterministic description}

To study the  dynamics of this model in the continuum limit, the effective population size
is assumed to be large, i.e. $N_e\gg 1$, and the allele frequencies are treated as continuous variables.
Denoting by $x$ the frequency of the mutant allele  $\textsf{A}'$ (the frequency of  $\textsf{A}$ is thus $y\equiv 1-x$)
the setting can be  summarized by the following the table:
\begin{center}
\begin{tabular}{c|c  c  c }
Genotypes & $\textsf{A}\textsf{A}$  & $\textsf{A}'\textsf{A}'$ & $\textsf{A}'\textsf{A}$ \\
 \hline
Relative fitness & $1$ & $1-s$ & $1-h$ \\
Frequency &  $(1-x)^2$ & $x^2$ & $x(1-x)$ \\
\end{tabular}
\end{center}
Accordingly, the  population average fitness is
\begin{eqnarray}
\label{wbar}
\bar{w}(x)=1-sx[x+2(h/s)(1-x)].
\end{eqnarray}
The change in the frequency of the allele  $\textsf{A}'$ in one generation $\delta \tilde{t}$ caused by random
mating
 is given by
$\delta x(\tilde{t}) \equiv x(\tilde{t}+\delta \tilde{t}) - x(\tilde{t})=
x(\tilde{t})\left[-1 + \left\{(1-s)x(\tilde{t}) +(1-h)(1-x(\tilde{t}))\right\}/\bar{w}(\tilde{t})\right]$~(\cite{Ewens,CrowKimura,Kimura1980})~\footnote{It is worth noticing that $s$ and $h$ are often assumed to be {\it vanishingly small} and the population average fitness is approximated to
be $\bar{w}(x)\approx 1$. Here, we {\it do not} make this simplifying assumption (whose practical validity is difficult to assess).}.
 Furthermore, at each generation the irreversible mutation $\textsf{A} \to
\textsf{A}'$  causes a change
$\delta x(\tilde{t})= v(1-x(\tilde{t}))$ in the   $\textsf{A}'$ allele frequency.
By  putting $\delta \tilde{t}=(2N_e)^{-1}$ we set the timescale in units of  $t\equiv\tilde{t}/(\delta \tilde{t})=2N_e\tilde{t}$, so that
 each allele has, on average, interacted once per time unit.
As  $\delta x= dx/dt + {\cal O}(N_e^{-1})$, when $N_e\gg 1$ and all fluctuations are neglected, the allele frequency varies according to the
rate equation (RE):
\begin{eqnarray}
\label{RE}
\frac{dx}{dt}=-\frac{x(1-x)}{\bar{w}(x)}[sx+h(1-2x)]+v(1-x).
\end{eqnarray}
This rate equation  is
characterized by the absorbing fixed point $x_{A'}^*=1$,
corresponding to a homozygous mutant  population ($\textsf{A}'\textsf{A}'$).
In addition, we shall see that in the biologically relevant setting where $v\ll s$,
there is always an  {\it attracting} fixed point
$0<x^*<1$ associated with a polymorphic (heterozygous) population ($\textsf{A}\textsf{A}'$).
In fact, in the sequel we will discuss in detail the following two main scenarios: when
$\textsf{A}'$ is completely dominant ($s=h>0$), $x_{A'}^*$ and $x^*$ are both attracting and separated
by a repelling fixed point $x_u^*$. Whereas, when the mutant allele is either recessive ($h=0, s>0$) or semi-dominant
($h=s/2>0$), $x^*$ is the only  attractor and $x_{A'}^*$ is the sole  repellor. Below, the latter will be referred to as scenario A and the former will be called scenario B.

\subsection{The stochastic description}
While the deterministic equation (\ref{RE}) aptly describes the variation of allele frequency
when $N_e \to \infty$, it does not account for fluctuations and stochastic effects that are of great importance
in a finite population ($N_e < \infty$). In particular, these fluctuations are responsible for the
{\it unavoidable} fixation of the mutant allele $\textsf{A}'$. The description of such a phenomenon therefore requires
to adopt a stochastic formulation of the evolutionary  dynamics.
A widely used stochastic model of evolving populations with non-overlapping generations was introduced by Fisher and Wright~(\cite{Fisher,Wright31}). A closely related and influential model for populations with overlapping generations,
that can be formulated as a Markov chain~(\cite{Feller,Ewens}),
was then introduced by Moran~(\cite{Moran58,Moran62}).
 The Moran model (MM) has the advantage of being mathematically more  amenable than the Wright-Fisher model (WFM)
and, when the population size is large, shares its  properties~(\cite{Ewens,Etheridge,Blythe})~\footnote{\label{foot5}In the neutral case (no selection and no mutations), a generation
in the WFM is $ {\cal N}/2$ times as long as in the MM, where $ {\cal N}$ is the total
number of individuals involved in the dynamics (here, ${\cal N}=2N_e$). That is, the
timescales of the WFM and MM differ by a factor ${\cal N}/2$~(\cite{CrowKimura,Ewens,Blythe}).}.
The MM was  originally formulated for haploid organisms, but has also been recently used for diploid populations (\cite{Durrett,Eriksson}).
In this work,  we will consider a continuous-time version of the MM with fitness-dependent transition rates [defined by
Eqs.~(\ref{fit1})-(\ref{trans})].

The dynamic properties of
continuous-time birth-death processes like the MM are suitably described by the following master equation that
gives the  probability distribution function (PDF) $P_n(t)$ of finding $n$ mutant alleles $\textsf{A}'$ at time $t$ in
the population~(\cite{Gardiner,VanKampen}):
\begin{eqnarray}\label{master}
\frac{d P_n(t)}{dt}& =& T^+_{n-1} P_{n-1}(t)+T^-_{n+1} P_{n+1}(t)-[T^+_{n}+T^-_{n}]P_n(t).
\end{eqnarray}
Here, $T^+_{n}$ and $T^-_{n}$ respectively denote the transition rates from a population state comprising $n$
mutants $\textsf{A}'$ to another state respectively with $n+1$ and $n-1$ mutant alleles $\textsf{A}'$ (see below).
For notational convenience, it is useful to denote the effective number of alleles by ${\cal N}\equiv 2N_e$.
Here, as the state space is bounded, with $n\in[0,{\cal N}]$, and $n={\cal N}$ is an absorbing state (all $\textsf{A}'$'s),
the transition rates satisfy $T^{\pm}_{{\cal N}}=0$, and it is assumed that $P_{n}(t)=0$ when $n<0$ and $n>{\cal N}$. With $n={\cal N}$ being absorbing, the stochastic dynamics eventually leads to the fixation of the mutant allele $\textsf{A}'$~(\cite{CrowKimura,
KimuraOhta,Ewens}).

In the continuous-time MM, the population evolves by pairs of alleles being sampled uniformly with replacement
from the population (the number of birth/death events follows a Poisson process
and the  waiting time is thus exponentially distributed). One of the sampled alleles is designated to be the parent and is
copied, yielding an offspring (birth) that replaces the other allele that is sacrificed (death). Thus, in the fitness-dependent
version of the MM that we consider, at each time step and after sampling,
the  pairs $\textsf{A}' \textsf{A}$ can  become either $\textsf{A}' \textsf{A}'$ (birth of $\textsf{A}'$, $n\to n+1$) with probability $p_{\textsf{A}'}$, or  $\textsf{A} \textsf{A}$ (birth of $\textsf{A}$, $n\to n-1$) with probability $p_{\textsf{A}}$.
These probabilities are  expressed in terms of the reproductive potential, or relative
fitnesses, of type $\textsf{A}'$ and $\textsf{A}$ respectively denoted $f_{\textsf{A}'}(n)$ and $f_{\textsf{A}}(n)$, i.e.
$p_{\textsf{A}'}\equiv (n/{\cal N})f_{\textsf{A}'}(n)/{\bar{w}}(n)$ and
$p_{\textsf{A}}\equiv (1-(n/{\cal N}))f_{\textsf{A}}(n)/{\bar{w}}(n)$~(\cite{CrowKimura}).
Here,  ${\bar{w}}(n)\equiv (1-(n/{\cal N}))f_{\textsf{A}}(n)+(n/{\cal N})f_{\textsf{A}'}(n)$ is the population mean genotypic fitness
and,
according to the table leading to (\ref{RE}), one has:
 \begin{eqnarray}\label{fit1}
 f_{\textsf{A}'}(n)&=& (1-s)\left(\frac{n}{{\cal N}}\right) + (1-h)\left(1-\frac{n}{{\cal N}}\right) , \nonumber\\
f_{\textsf{A}}(n)&=&  \left(1-\frac{n}{{\cal N}}\right)+(1-h)\left(\frac{n}{{\cal N}}\right).
\end{eqnarray}

 In addition to birth and death events, we also have to account for the $\textsf{A} \to \textsf{A}'$ mutations, which can be implemented in various ways
within the MM~(see, e.g., \cite{CrowKimura,Ewens,Blythe}).
To ensure a neat connection with the deterministic rate equations (\ref{RE}), that we wish to recover in the limit ${\cal N} \to \infty$ (see below), our approach here, as in (\cite{Blythe}), is to consider
the mutation process as being divorced from the death-birth events. Hence, death/reproduction and mutation  take place independently, with mutations occurring spontaneously (not necessarily between a death and a birth). Thus, on average, at each time step each pair of alleles undergoes the above birth-death process and
is sampled as described above with probability $1-v$. The rest of the time, with probability $v$, the  $\textsf{A}$ offspring (if any) is switched into the allelic type $\textsf{A}'$.
Therefore, following the above discussion and using~(\ref{fit1}), the transition rates $T^+_{n}$ and $T^-_{n}$ of the birth $\left(n\textsf{A}' \to (n+1)\textsf{A}'\right)$ and death $\left(n\textsf{A}' \to (n-1)\textsf{A}'\right)$ processes appearing in~(\ref{master}), read:
 \begin{eqnarray}\label{trans}
T^+_{n}&=& \left(1-\frac{n}{{\cal N}}\right)\, \left[(1-v) p_{\textsf{A}'} + v\right]=\left(1-\frac{n}{{\cal N}}\right)\left[\left(\frac{n}{{\cal N}} \right)\frac{(1-v)f_{\textsf{A}'}(n)}{\left(\frac{n}{{\cal N}}\right)f_{\textsf{A}'}(n)+\left(1-\frac{n}{{\cal N}}\right)f_{\textsf{A}}(n)}+ v\right]\nonumber\\
T^-_{n}&=&\left(\frac{n}{{\cal N}}\right)\, (1-v)p_{\textsf{A}}=\left(\frac{n}{{\cal N}}\right)\left(1-\frac{n}{{\cal N}}\right)\left[\frac{(1-v) f_{\textsf{A}}(n)}{\left(\frac{n}{{\cal N}}\right)f_{\textsf{A}'}(n)+\left(1-\frac{n}{{\cal N}}\right)f_{\textsf{A}}(n)}\right].
\end{eqnarray}
We notice that if time is measured in units of $(1-v)t$,  the
factors $1-v$ appearing in the expressions of $T^{\pm}_n$ are eliminated, while  the mutation rate on the right-hand-side of $T^{+}_n$
becomes $v/(1-v)$. As we are interested in the limit of small mutation rates and neglect terms of order ${\cal O}(v^2)$, throughout this work
we shall simply  consider  $v/(1-v)= v+{\cal O}(v^2)$ and measure time in units of $(1-v)t\simeq t$.
In the continuum limit (where ${\cal N} \gg 1$), the RE describing the average number $\bar{n}$ of mutant alleles $\textsf{A}'$ can be directly obtained from Eq.~(\ref{master}). It reads $(d/dt)\bar{n}=T^+_{n}-T^-_{n}$, where upon rescaling time as in Sec. 2.1~\footnote{The time appearing in (\ref{master}) is rescaled according to $t\to {\cal N}t$. As in  Sec. 2.1, this means that the time step is $\delta t={\cal N}^{-1}$ and therefore, on average, each allele is sampled once per time unit.}, one recovers Eq.~(\ref{RE}) for the frequency $x(t)=\bar{n}(t)/{\cal N}$ of mutants.

While it is generally a very demanding task to extract accurate and useful information from Eq.~(\ref{master}), the latter can be investigated within various approaches. One very popular and insightful approximation is provided by the diffusion theory~(see, e.g., \cite{CrowKimura,Kimura,Ewens}), that is based on a Taylor-expansion in ${\cal N}^{-1}\ll 1$ of the master equation leading to a  Fokker-Planck or Kolomogorov equation (KE)~(\cite{Gardiner,VanKampen,Risken}). For the problem at hand, the backward KE associated with the birth-death process~(\ref{master}) for the probability density ${\cal P}(x,t)\equiv {\cal N} \,P_n(t)$, see below, is given by
\begin{eqnarray}
 \label{bKE}
\frac{\partial {\cal P}(x,t)}{\partial t}&=& {\cal L}_{{\rm bKE}}(x,t) \,{\cal P}(x,t), \quad \text{with}\nonumber\\
{\cal L}_{{\rm bKE}}(x)&\equiv& M(x)\frac{\partial}{\partial x}+ \frac{1}{2}V(x)\frac{\partial^2 }{\partial x^2}.
\end{eqnarray}
In this equation,
$M(x)$ and $V(x)$ represent the deterministic drift and
the diffusion terms, respectively. The connection
between (\ref{bKE}) and  the evolutionary dynamics is made by determining the functions $M(x)$ and $V(x)$
from the original (non-approximate) stochastic processes. For this, one notes that according to (\ref{bKE}) the mean
of the allele frequency $x$ and the average of its square obey the equations of motion (EOMs) $(d/dt)\langle x(t)
\rangle\equiv  \int dx \, x {\cal P}(x,t)=\langle M(x) \rangle$ and $(d/dt)\langle x^2(t)
\rangle\equiv  \int dx  \, x^2 {\cal P}(x,t)=2 \langle x M(x) \rangle + \langle V(x) \rangle$. The functions $M$ and $V$ are then determined
by imposing consistency with the EOMs of $\langle x(t) \rangle$ and $\langle x^2(t) \rangle$ derived from the
  original stochastic processes, here defined by (\ref{master})-(\ref{trans}), which yields  $(d/dt)\langle x(t) \rangle\equiv \sum_{n} (n/{\cal N}) (dP_n(t)/dt)$
 and $(d/dt)\langle x^2(t) \rangle\equiv \sum_{n} (n/{\cal N})^2 (dP_n(t)/dt)$.
Therefore, for the dynamics based on the MM~(\ref{fit1})-(\ref{trans}), after rescaling time according to $t\to {\cal N}t$
(see Footnote 7),  the ensuing
 EOMs for $\langle x(t) \rangle$ and $\langle x^2(t) \rangle$ give
 $M_{{\rm MM}}(x)\equiv T_n^+-T_n^-$ and $V_{{\rm MM}}(x)\equiv(T_n^+ +  T_n^-)/{\cal N}$. Thus,
for the model that we consider here
in the limit $v\ll s,h\ll 1$ and ${\cal N}\gg 1$, one finds
$M_{{\rm MM}}(x)\approx x(1-x)[sx+h(1-2x)]+v(1-x)$ and $V_{{\rm MM}}(x)\approx 2x(1-x)/{\cal N}$.

In
the analysis of Ref.~(\cite{Kimura}), the  ``microscopic'' dynamics was
implemented according to the WFM based on a discrete binomial sampling process defined
by a transition matrix $[a_{ij}]$, with $i,j=0,1, ..., {\cal N}$ and $a_{ij}=\{{\cal N}!/[j!({\cal N}-j)!]\}\left(Q_i\right)^j
(1-Q_i)^{{\cal N}-j}$, where $Q_i=q_i+q_i(1-q_i)[sq_i+h(1-2q_i)]/{\bar{w}}$ and $q_i=(i/{\cal N})+v[1-(i/{\cal N})]$.
In this case, Kimura showed~(\cite{Kimura}) that $M_{{\rm WFM}}$ and $V_{{\rm WFM}}$  are respectively the mean and the variance of the binomial distribution and therefore,
in the limit $v\ll s,h\ll 1$ and ${\cal N}\gg 1$ with $x\equiv i/{\cal N}$,
 $M_{{\rm WFM}}(x)\approx x(1-x)[sx+h(1-2x)]+v(1-x)$ and $V_{{\rm WFM}}(x)\approx x(1-x)/{\cal N}$ (adopting the same timescale as in Sec.~2.1).
Kimura, then used these expressions in the differential operator ${\cal L}_{{\rm bKE}}(x,t)$ of Eq.~(\ref{bKE}) to obtain the MFT from the diffusion
approximation. We thus notice that
 the expressions of
$M(x)$ coincide for the WFM and MM (i.e. $M_{{\rm MM}}(x)=M_{{\rm WFM}}(x)=M(x)$), but the function $V(x)$ obtained for the
MM has a factor $2$ compared to that of the WFM, i.e. $V_{{\rm MM}}(x)=2V_{{\rm WFM}}(x)=2V(x)$. This difference
stems from the fact that (in the neutral case) a generation
in the WFM is $ {\cal N}/2$ times as long as in the MM (see Footnote~\ref{foot5})
and means that the results for the MM can be mapped onto those of
WFM (within the realm of the diffusion approximation) via the transformation ${\cal N}=2N_e \to {\cal N}/2$.
In other words, the predictions of the  MM comprising ${\cal N}$ (breeding) alleles
should be compared with those of the WFM made up of
 ${\cal N}/2$ (breeding) alleles.

In this work, our goal is to accurately compute the MFT and QSD of the mutant allele $\textsf{A}'$ for the MM defined by (\ref{master})-(\ref{trans}), beyond the limit of validity of the diffusion theory. Here, when $s$ is finite, fixation is triggered by a rare large fluctuation and the non-diffusive
dynamics is  characterized by a QSD with \textit{non-Gaussian (``fat'') tails}. Therefore, we will show that this phenomenon cannot be accurately described by  Eq.~(\ref{bKE}) in the realm of the diffusion approximation.
It turns out that the eigenvectors and eigenvalues associated with the transition matrix of the underlying birth-death process
(\ref{master})-(\ref{trans}) can be obtained analytically and be used to formally compute the MFT and QSD~(\cite{VanKampen,Ewens,Gardiner}).
However, while these results are exact, they are non-generic (i.e., limited to diallelic models), and it is very difficult to extract their asymptotic behavior.
It is therefore desirable to develop reliable and more generally applicable approximation methods to study stochastic evolutionary problems~(\cite{MA,MAMM}).

Here, we employ a WKB (Wentzel-Kramers-Brillouin) theory~(\cite{Landau}),
to analyze the properties of the master equation~(\ref{master})~(\cite{kubo,Dykman}).
The WKB approximation, sometimes referred to as semi-classical or eikonal approximation, is an asymptotic  theory
frequently used in the semi-classical treatment of quantum mechanics.
This method has been recently used in the
context of population dynamics where it allowed to accurately study extinction/fixation from a metastable state as the
result of large fluctuations,
see, e.g., (\cite{AM,MA,MAMM}) and references therein.
Our WKB-based analytical treatment is supported by numerical solutions of the master equation and Monte Carlo (MC) simulations, whose implementation is briefly described in the Appendix.

\begin{figure}
\begin{center}
\includegraphics[width=4.5in,height=2.75in,clip=]{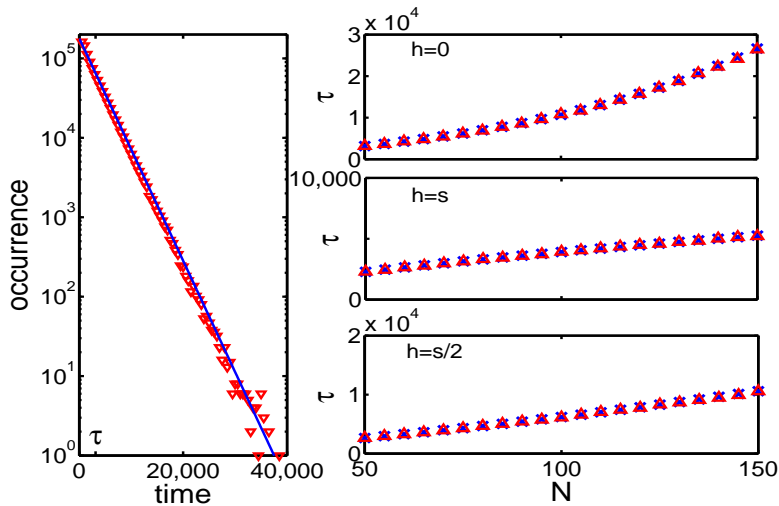}
\caption{On the left shown is a histogram of fixation times (triangles) obtained from MC simulations with  $10^6$ runs, for ${\cal N}=50$, $s=0.1$, $h=0$, and $v=0.04$ for the dynamics according to the MM. The fixation times are found to be exponentially distributed as indicated by the excellent agreement with an exponential distribution (solid line) with mean $\tau=3,158$.
On the right shown are the mean fixation times $\tau$
 versus the (effective) population size ${\cal N}$
for the cases of recessive (upper panel), completely dominant (middle panel), and semi-dominant mutants (bottom panel). Here $s=0.1$ and $v=0.04$.
The numerical solution of the master equation~(\ref{master}) ($\times$'s) agree excellently with results of MC simulations (triangles) averaged over $100,000$ runs.}
\label{MCvsExact}
\end{center}
\end{figure}

While the crux of the WKB method is given in the next section, let us first gain some insight into the problem at hand, by looking at numerical results reported in Fig.~\ref{MCvsExact}. On the left panel of Fig.~\ref{MCvsExact}, we report a histogram of fixation times obtained from a MC simulation with $10^6$ runs, see Appendix. The latter is excellently fitted by an exponential distribution, which is a characteristic feature of systems displaying metastability
and where fixation/extinction is driven by  large fluctuations
(see \textit{e.g.} \cite{AM1,AM2})). The average of this fixation time distribution is the MFT, denoted by $\tau$,
and is the quantity that is computed in the next sections. On the right panels of Fig.~\ref{MCvsExact} we compare the MFT obtained from MC simulations (averaged over $100,000$ runs) and by numerically solving the master equation~(\ref{master}), and observe an excellent agreement. This indicates that averaging over sufficient MC runs accurately reproduces the stochastic dynamics of the system described by the master equation (\ref{master}).

\section{The WKB approach}
In this section, the main general aspects of the  WKB approach to treat the birth-death process (\ref{master})-(\ref{trans}) are presented.
The basic idea relies on the fact that in the presence of demographic fluctuations the process is
characterized by {\it metastability}, with the frequency of $\textsf{A}'$ lingering around the metastable value $x^*$. After a very long time of average $\tau$, the mutant allele  $\textsf{A}'$ eventually  takes over and fixates the entire population
due to the combined effect of selection, weak (yet steady) mutation pressure and random fluctuations.

To proceed analytically the following key assumptions are made: (i) The population size is large (but finite), i.e.
${\cal N}\gg 1$. (ii) Fixation {\it always} occurs from the interior fixed point $x^*$ that is reached after a typical relaxation time $t_r\ll \tau$. That is, we assume the system converges into the vicinity of the fixed point \textit{prior} to fixation~\footnote{This assumption is always satisfied here since it is natural to assume that only a few mutant alleles are initially present. In any case, we assume that the initial number of mutants is not too close to ${\cal N}$, in which case fixation does not occur instantaneously.}.

The crux of the analytical treatment of the master equation is an expansion of
$P_n(t)$, the system's PDF, into a series
of eigenvectors and eigenvalues of the stochastic (Markov) generator, (see \textit{e.g.} \cite{AM0,AM} and references therein). After a
relaxation time $t_r$  the system settles  in the metastable state, where the frequency of alleles $\textsf{A}'$
is (sharply) distributed around $x^*$, for a very long time. Such a metastable state is encoded in the first excited eigenvector,
$\pi_n$, of the Markov chain (\ref{master}) that has not vanished after $t\geq t_r$, (see \textit{e.g.} \cite{AM1,AM2}). The quantity
$\pi_n$ therefore determines the QSD (the shape of the metastable PDF), while the decay rate of the metastable
distribution is given by the first nonzero eigenvalue of the Markov chain~(\ref{master}),
which is the inverse of the MFT. Hence, at $t\gg t_r$, the slow decay of the metastable PDF $P_{n<{\cal N}}$ and
the slow increase of the fixation probability $P_{{\cal N}}$ are given
 by~(\cite{AM1,AM2})
\begin{eqnarray}\label{qsd}
P_{0\leq n<{\cal N}}(t)\simeq \pi_n e^{-t/\tau} \; \quad \text{and} \quad \;P_{{\cal N}}(t)\simeq 1-e^{-t/\tau}\,,
\end{eqnarray}
where $\tau$, the mean decay time of the metastable state, is the mean time it takes the system to fixate. It follows from (\ref{master}) and (\ref{qsd}) that $(d/dt)P_{{\cal N}}=(1/\tau)e^{-t/\tau}=T^{+}_{{\cal N}-1}\,\pi_{{\cal N}-1}\,e^{-t/\tau}$,
which readily gives
\begin{equation}
\label{tau}
\tau=[T^+_{{\cal N}-1}\pi_{{\cal N}-1}]^{-1}.
\end{equation}
This equation simply means that the mean fixation rate is given by the \textit{flux} of probability
into the absorbing state $n={\cal N}$. To obtain such a quantity, we need to compute $\pi_{{\cal N}-1}$. It is also valuable
to compute the entire QSD that is characterized by
markedly non-Gaussian (``fat'') tails. Note, that the expression (\ref{tau}) is
independent of the number of mutants alleles $n$ initially present in the population. This reflects the assumption that fixation
always occurs from the metastable state that serves here as an effective initial condition, see Footnote 8.

To determine the QSD we now assume that the MFT is exponentially large in ${\cal N}\gg 1$ (which will be checked a posteriori)
and substitute the expressions (\ref{qsd}) into (\ref{master}). Upon neglecting the  exponentially small term $\pi_n/\tau$, one thus finds
that the QSD obeys the following quasi-stationary master equation:
\begin{eqnarray}\label{masterqsd}
0= T^+_{n-1} \pi_{n-1}+T^-_{n+1} \pi_{n+1}-[T^+_{n}+T^-_{n}]\pi_n.
\end{eqnarray}

We will now solve this equation using the WKB approximation. Originally, the WKB approximation  was introduced  to
treat ordinary differential equations where the highest-order derivative is multiplied by a small parameter~(\cite{Landau}).
Here, with $x\equiv n/{\cal N}$ regarded as a continuous variable when ${\cal N}\gg 1$, we employ the following WKB ansatz (i.e. a trial function) [see, e.g.,~(\cite{kubo,Dykman})]
\begin{equation}
\label{ansatz}
\pi_n\equiv \pi_{{\cal N}x}=\pi(x)\simeq {\cal A} e^{-{\cal N}S(x)-S_1(x)-\dots},
\end{equation}
where $S(x), S_1(x), \ldots$ are all assumed to be of order unity.
Substituting ansatz~(\ref{ansatz}) into Eq.~(\ref{masterqsd}) yields in the leading order in ${\cal N}\gg 1$ an equation for $S(x)$, whose solution may be used to solve the subleading-order equation for $S_1(x)$, etc. In analogy with dynamical systems~(\cite{Landau}), $S(x)$ is referred to as the action while $S_1(x)$ is called the amplitude.
 Here, ${\cal A}$ is a constant introduced here for convenience.
Note that  higher-order corrections in the exponent of (\ref{ansatz}) are of order ${\cal O}({\cal N}^{-1})$ and hence negligible.

Finding the MFT of the mutant species $\textsf{A}'$ can be rephrased as determining the mean time to extinction (MTE)
of the wild type  $\textsf{A}$ in the aftermath of a long-lived metastable coexistence of both species.
The problem of finding the MTE in a generic two-state system has recently been studied in Ref.~(\cite{AM}) whose results can be used here.
In the following, for the sake of clarity, we briefly repeat the leading order calculation of the MFT (or MTE) and QSD and then outline how subleading-order corrections are obtained.

The leading-order calculations require finding the action $S(x)$. Substituting ansatz~(\ref{ansatz}) into Eq.~(\ref{masterqsd}), and using the continuous counterpart of the transition rates~(\ref{trans}),
${\cal T}_{\pm}(x)=T^{\pm}(n/{\cal N})$, we obtain in the leading  order (\cite{Dykman})
\begin{eqnarray}
\label{eqS}
0={\cal T}_+(x) \left(e^{S'(x)}-1\right)+{\cal T}_-(x) \left(e^{-S'(x)}-1\right),
\end{eqnarray}
whose solution is
\begin{equation}\label{Ssingle}
S(x)=-\int^x\ln [{\cal T}_{+}(\xi)/{\cal T}_-(\xi)]\,d\xi.
\end{equation}
This corresponds to an integral over what is called the ``optimal path to extinction".
The latter refers to the path followed by the stochastic system that leaves
 the metastable state at $t=-\infty$ and arrives at the absorbing state at $t=\infty$ (see \cite{Dykman} for a detailed discussion).

To relate the MFT and the action, we  write in the continuum limit $T^+_{{\cal N}-1}\simeq |{\cal T}_+'(1)|/{\cal N}$ [since ${\cal T}_+(1)$ vanishes] and $\pi_{{\cal N}-1}\simeq \pi(1)$, and thus,  expression (\ref{tau})  can be rewritten as
\begin{equation}
\label{tau1}
\tau=\frac{{\cal N}}{|{\cal T}_+'(1)|\pi(1)}.
\end{equation}
Furthermore, we have seen that the QSD is peaked in the vicinity of $x^*$, where the relative
 width of the distribution scales as ${\cal N}^{-1/2}$. Therefore, the QSD is strongly peaked around $x^*$ when  ${\cal N}\gg 1$,
and the constant ${\cal A}$ can be found by approximating the QSD by a Gaussian centered at $x^*$ and normalized to unity (\cite{EsK,AM}).
Indeed, for $x\simeq x^*$ one can write $\pi(x)\simeq {\cal A}
e^{-{\cal N}S(x^*)-S_1(x^*)-({\cal N}/2)S''(x^*)(x-x^*)^2}$, whose normalization yields
in the leading order ${\cal A}\sim e^{{\cal N}S(x^*)}$. Therefore, to leading order one has
\begin{equation}\label{pinorm}
\pi(x)\sim e^{-{\cal N}[S(x)-S(x^*)]}.
\end{equation}
As a result, using Eq.~(\ref{tau1}), one finds
\begin{equation}\label{mainMFT}
\tau\sim  e^{{\cal N}\Delta S},
\end{equation}
where $\Delta S=S(1)-S(x^*)$ is the ``accumulated action'' over the optimal path to extinction.
It is worth noticing that in the case of complete dominance the absorbing and interior fixed points, $x_{A'}^*$ and $x^*$,
are separated by a repelling fixed point $x_u^*$. Therefore, in this case the  accumulated action strictly reads
$\Delta S=S(x_u^*)-S(x^*)$. However, in the limit of weak mutation rate ($v \ll s$) that is of interest to us,
one has $x_u^*\simeq 1$. For the three cases of interest, we thus have
$\Delta S\simeq S(1)-S(x^*)$ (\cite{EsK,AM}).

Expressions (\ref{pinorm}) and (\ref{mainMFT}) therefore give the generic leading-order results for the QSD and the MFT of
the problem, whereas the action function $S(x)$ has to be calculated separately for
the cases of completely dominant, semi-dominant
and recessive mutant alleles (see Section 4). The ensuing results are valid provided
  that ${\cal N}\Delta S\gg 1$ (for the MFT to be exponentially large as required by the WKB ansatz), along with the necessary condition  ${\cal N}\gg 1$.

The calculation of the subleading corrections to the MFT is more involved.
It has in fact been shown that for a generic two-state problem extinction (and therefore fixation) occurs via two scenarios, called scenarios
 $A$ and $B$ in Ref.~(\cite{AM}). Here, the cases of a semi-dominant and completely recessive mutant allele, where  the absorbing ($x_{A'}^*$) and interior ($x^*$) fixed points of (\ref{RE}) are respectively repelling and attracting, correspond to scenario A.
On the other hand, the case of a completely dominant mutant allele where both $x_{A'}^*$ and $x^*$ are attracting and separated by a repelling fixed point, corresponds to scenario B.
We now quote the results and outline the main steps to the subleading-order calculations and refer to Ref. (\cite{AM}) for technical details.
To determine the subleading-order contribution to the QSD and MFT in problems belonging to scenario A,
the WKB solution~(\ref{pinorm}) needs to be matched with a  recursion solution of the (quasi-stationary) master equation~(\ref{masterqsd}) in the close vicinity of the
absorbing state $n={\cal N}$. This is because the WKB solution~(\ref{pinorm}) is only valid sufficiently far from the absorbing state.
In Scenario B, the WKB solution already breaks down in the vicinity of the
intermediate repelling fixed point $x_u^*$. Thus, the master equation~(\ref{masterqsd}) has to be solved
 in the a close vicinity of
$x_u^*$ (by using, \textit{e.g.}, a Fokker-Planck approximation). This solution needs to be matched  on the one hand with the
 WKB solution for $x<x_u^*$, and on the other hand with the recursion solution of Eq.~(\ref{masterqsd}) when $x>x_u^*$ (\cite{AM}).

Implementing these steps and using the action (\ref{Ssingle}), one finds for the QSD~(\cite{AM})
\begin{equation}\label{qsdfull}
\pi(x)=\frac{\sqrt{S''(x^*)}{\cal T}_+(x^*)}{\sqrt{2\pi{\cal N}{\cal T}_+(x){\cal T}_-(x)}}e^{-{\cal N}[S(x)-S(x^*)]}\,,
\end{equation}
while the MFT in Scenarios A  and B  is respectively given by $\tau^{\rm {SA}}$ and $\tau^{\rm {SB}}$, where
\begin{equation}\label{MFTfull}
\tau^{\rm {SA}}=\frac{\sqrt{2\pi {\cal N} R}}{{\cal T}_+(x^*)(R-1)\sqrt{ S''(x^*)}}e^{{\cal N}[S(1)-S(x^*)]}
\;,\;\;\;\;\tau^{\rm {SB}}=\frac{2\pi{\cal N}}{{\cal T}_+(x^*)\sqrt{S''(x^*)|S''(x_u^*)|}}e^{{\cal N}[S(x_u^*)-S(x^*)]},
\end{equation}
with $R={\cal T}_-'(1)/{\cal T}_+'(1)$. These results hold for
two-state populations undergoing generic single-step processes~\footnote{In the case where the rescaled transition
rates ${\cal T}_{\pm}(x)$ also include subleading order corrections of order ${\cal O}({\cal N}^{-1})$, an additional ${\cal O}(1)$ prefactor enters the final results~(\ref{qsdfull}) and (\ref{MFTfull}), see Ref.~(\cite{AM}).}. It is worth noticing that  the subleading prefactors in
(\ref{qsdfull}) and (\ref{MFTfull}) scale as some power of ${\cal N}\gg 1$.

\section{Results for the MFT and QSD}
In this section we present our results  for the MFT and QSD in the cases of (i) complete dominance, (ii) recessivity and (iii)
semi-dominance and focus on the limit where the mutation rate is much smaller than the selection
intensity, i.e. $v\ll s\ll 1$, see Footnote~4.
To ensure the validity of the WKB treatment, we
also assume  that $s\gg {\cal N}^{-1}$, i.e. the selection intensity $s$ is ``not too small'' (see below).
For all cases (i)-(iii)  we shall give the detailed analytical derivation of the action
 (\ref{Ssingle}) to order ${\cal O}(v)$
that yields the leading contribution to the QSD and MFT according to~(\ref{pinorm}) and (\ref{mainMFT}). We also check our findings against
results of numerical simulations and report the analytical results for
the MFT and QSD that include pre-exponential contributions.

In the continuum variable $x=n/{\cal N}$, the transition rates in our problem are given
by
\begin{eqnarray}
 \label{tau-x}
{\cal T}_{+}(x)=
\frac{x(1-x)[1-h+(h-s)x]}{1-sx[x+2(h/s)(1-x)]} + v(1-x)\quad \mbox{and} \quad
{\cal T}_{-}(x)=
\frac{x(1-x)[1-hx]}{1-sx[x+2(h/s)(1-x)]}.
\end{eqnarray}

For convenience, in the rest of this section we will work in terms of the variable $y=1-x$ corresponding to the frequency of the wild type
 $\textsf{A}$. In terms of the $y$ variable,  the absorbing state
corresponds to $y=0$. Also, under the mapping  $x\to y=1-x$
the transition rates are transformed into ${\cal T}_{\pm}(x)\to {\cal T}_{\mp}(y)$.

\subsection{Results in the case of complete dominance $(s=h>0)$}
When the mutant allele $\textsf{A}'$ is completely dominant, i.e. $s=h>0$,
the rate equation (\ref{RE}) admits the following interior fixed points ($y=1-x$)
\begin{eqnarray}
 \label{fp-dom}
y^{*}_{\pm}=  \frac{1 \pm \sqrt{(1+2v)^2-4v(1+v)/s}}{2(1+v)}.
\end{eqnarray}
As we consider $v\ll s\ll 1$, these expressions can be simplified,
yielding $y_+^* =1-v/s + {\cal O}(v^2)$ and  $y_-^* = (1-s)v/s
+ {\cal O}(v^2)$. It can be readily checked that $y_+^*$ and $y_-^*$ are respectively attracting and repelling.
 Adopting the notation of Sec.~2, we can write $y^*=y_+^*$ and  $y_u^*=y_-^*$, and in this case the metastable state $y^*$ and the absorbing state $y_{{\textsf A}'}=0$ are
 separated by a repelling fixed point  $y_u^*$, which indicates that this problem corresponds to scenario B of extinction. In this case, the transition rates (\ref{tau-x}) read:
\begin{eqnarray}
 \label{rates-dom}
{\cal T}_{+}(y)=\frac{(1-y) y [s (y-1)+1]}{1-s \left(1-y^2\right)} \quad \mbox{and} \quad
{\cal T}_{-}(y)=\frac{y [s (y-1) (v y+v+1)+v-y+1]}{1-s \left(1-y^2\right)}.
\end{eqnarray}
Thus, $$\frac{{\cal T}_{-}(x)}{{\cal T}_{+}(y)}= \frac{1-s}{1-s (1-y)}+\frac{1-s(1-y^2)}{(1-y) [1-s (1-y)]}\,v.$$
The main contribution to the MFT and QSD is determined by the action~(\ref{Ssingle}). To leading  ${\cal O}(v)$ order, we find
\begin{equation}\label{S2}
S(y)=-\int^y\ln [{\cal T}_{+}(\xi)/{\cal T}_-(\xi)]\,d\xi\simeq y-\left(\frac{1-s}{s}+y\right)\ln\left(1+\frac{sy}{1-s}\right)+\frac{sy(2+y)+2\ln(1-y)}{2(1-s)}\,v+{\cal O}(v^2).
\end{equation}
Therefore, the accumulated action over the optimal path to extinction is given by
\begin{equation}
\label{DeltaS-dom}
\Delta S=S(y_u^*)-S(y^*)=
\frac{1}{s}\ln{\left(\frac{1}{1-s}\right)}-1+\,\frac{1}{1-s}\left[\ln\frac{v}{s}+\frac{s}{2(1+s)}+
\frac{3s}{2}+\frac{(1-s)\ln (1-s)}{2}\right]\,v+{\cal O}(v^2),
\end{equation}
while, from (\ref{S2}), to order ${\cal O}(v)$,
one finds $S''(y^*)\simeq s^2/v\;,\;S''(y_u^*)\simeq -s$.

Therefore, using Eqs~(\ref{qsdfull}) and (\ref{MFTfull}), the MFT and QSD in this case are explicitly given by
\begin{eqnarray}\label{MFT2}
\tau=\frac{2\pi {\cal N}}{\sqrt{sv}}e^{{\cal N} \Delta S},\;\;\;\pi(x)=\sqrt{\frac{v}{2\pi {\cal N}}}e^{-{\cal N}[S(x)-S(x^*)]}.
\end{eqnarray}
where $S(x)$ is given by Eq.~(\ref{S2}) with $x=1-y$.
Note, that while we have included here only
leading-order terms with respect to $s\ll 1$ in the prefactors, the exponent remains a nontrivial function of $s$,
see Eq.~(\ref{S2}). This is because even though ${\cal N}^{-1}\ll s\ll 1$, we are not making any assumptions
regarding the smallness of terms such as ${\cal N}s^{3/2}$, ${\cal N}s^2$, etc. In fact, as these
 terms are exponentiated in (\ref{DeltaS-dom}), all of them must be kept.

The validity of these results requires first of all that ${\cal N}\Delta S\gg 1$ (for the WKB theory to hold),
that is, $s \gg {\cal N}^{-1}$ (namely $s$ cannot be too small). Furthermore, since we have neglected terms of
order $v^2$ in $\Delta S$, $v$ must satisfy $v\ll {\cal N}^{-1/2}$. In addition, the theory assumes that $y^*$
and $y_u^*$ are sufficiently separated from the boundaries $y=1$ and $y=0$, respectively, with a Gaussian distribution in the close vicinity
of $y^*$, which also requires that $v\gg {\cal N}^{-1}\;$~\footnote{The requirement $v\gg {\cal N}^{-1}$ is only of technical convenience.
This condition does not stem from our theoretical approach; without such a condition, $y^*$ would not necessarily be sufficiently separated from  $y=1$
to allow a normalization by a Gaussian approximation, see text. Yet, up to a multiplicative ${\cal O}(1)$ factor,
one would obtain the same results as reported here.} (in addition to $s \ll 1$). Overall, the validity criteria can be summarized by
\begin{equation}\label{cond}
 {\cal N}^{-1}\ll v\ll {\cal N}^{-1/2}\;,\;\;v\ll s\ll 1\,.
\end{equation}
These conditions are generically fulfilled in the system that we consider in this work (see Footnote 4).

\subsection{Results in the case of complete recessivity $(h=0, s>0)$}
When the mutant allele $\textsf{A}'$ is completely recessive, i.e. $s>0$ and $h=0$,
the rate equation (\ref{RE}) admits a single attracting interior fixed point that reads

\begin{eqnarray}
 \label{fp-rec}
y^{*}=  1-\sqrt{\frac{v}{s(1+v)}}.
\end{eqnarray}
Again this term can be simplified for $v\ll s$ into  $y^*=1-\sqrt{v/s}+{\cal O}(v^{3/2})$. As explained above,
this case corresponds to scenario A of extinction. In this case, the transition rates (\ref{tau-x}) read:
\begin{eqnarray}
 \label{rates-rec}
{\cal T}_{+}(y)=\frac{(1-y) y}{1-s \left(1-y^2\right)}\quad \mbox{and} \quad
{\cal T}_{-}(y)=y\left[1+v-\frac{y}{1-s \left(1-y^2\right)}\right].
\end{eqnarray}
Therefore $$\frac{{\cal T}_{-}(x)}{{\cal T}_{+}(y)}= 1-s(1-y)+\left[\frac{1}{1-y}-s(1-y)\right]\,v$$
and, to order ${\cal O}(v)$, the action $S(y)$ given by Eq.~(\ref{Ssingle}) is
\begin{equation}\label{S1}
S(y)=-\int^y\ln [{\cal T}_{+}(\xi)/{\cal T}_-(\xi)]\,d\xi\simeq -y+\left(\frac{1-s}{s}+y\right)\ln [1-s(1-y)]
-\left\{1-y+\ln(1-y)+\frac{1-s}{s}\ln[1-s(1-y)]\right\}\,v.
\end{equation}
Up to order ${\cal O}(v^{2})$, the accumulated action between $y=0$ and $y=y^*$ is thus given by
\begin{equation}
\label{DeltaS1}
\Delta S=S(0)-S(y^*)=1+\frac{1-s}{s}\ln(1-s)+\left[\frac{1}{2}\ln \frac{v}{s}-\frac{3}{2}-\frac{1-s}{s}\ln(1-s)\right]\,
v+{\cal O}(v^{3/2}).
\end{equation}
Using these results, and the fact that $S''(y^*)\simeq 2s$, Eqs.~(\ref{qsdfull}) and (\ref{MFTfull}) become in this case
\begin{eqnarray}\label{MFT1-QSD1}
\tau=\frac{\sqrt{\pi{\cal N}}}{s\sqrt{v}}e^{{\cal N} \Delta S},\;\;\;\pi(x)=\sqrt{\frac{v}{\pi {\cal N}}}e^{-{\cal N}[S(x)-S(x^*)]}.
\end{eqnarray}
where $S(x)$ is given by Eq.~(\ref{S1}) with $x=1-y$ and $\Delta S$ by (\ref{DeltaS1}).
Here, similarly as in the completely-dominant case, the WKB results~(\ref{MFT1-QSD1}) are valid as long as condition~(\ref{cond}) holds (see Footnote 10).

\subsection{Results in the case of semi-dominant case $(h=s/2, s>0)$}
When the mutant allele $\textsf{A}'$ is semi-dominant, i.e. $s>0$ and $h=s/2$,
the rate equation also (\ref{RE}) again admits a single attracting interior fixed point that reads
\begin{eqnarray}
 \label{fp-semidom}
y^{*}=  1-\frac{2v}{s(1+2v)}.
\end{eqnarray}
Similarly as before, for $v\ll s$ this expression simplifies to $y^*=1-2v/s + {\cal O}(v^2)$. This case again corresponds to scenario A of extinction. Here, the transition rates (\ref{tau-x}) read:

\begin{eqnarray}
 \label{rates-semidom}
{\cal T}_{+}(y)=\frac{(1-y) y(2-s(1-y)}{1-s(1-y)}\quad \mbox{and} \quad
{\cal T}_{-}(y)=\frac{y\left[(1-y+v)-s(1-y/2+v)(1-y)\right]}{1-s(1-y)}.
\end{eqnarray}
Thus, the action function [Eq.~(\ref{Ssingle})] and accumulated action are here explicitly given by
\begin{equation}
\label{S-semi}
S(y)=\frac{2-s(1-y)}{s}\ln\left[1-\frac{s}{2-s(1-y)}\right]-\ln[2-s(2-y)]-\frac{2\{(1-s)\ln[2-s(2-y)]+\ln(1-y)\}}{2-s}v+{\cal O}(v^2)\,,
\end{equation}
\begin{eqnarray}
\label{DeltaS-semi}
&&\hspace{-10mm}\Delta S=S(0)-S(y^*)=\frac{2[(2-s)\ln \left(\frac{2(1-s)}{2-s}\right)-\ln (1-s)]}{s}\nonumber\\
&&\hspace{-8mm}+\frac{s^2[1-2(2-s)\ln(2-s)]-2(2-s)\left\{2\ln \left(\frac{2}{2-s}\right)-s\left[(s-2)\ln(2-2s)+\ln\left(\frac{8(1-s)v}{s}\right)\right]\right\}}{s(s-2)^2}v+{\cal O}(v^2)
\end{eqnarray}

\begin{figure}[ht]
\begin{center}
\includegraphics[width=4.0in,height=2.75in,clip=]{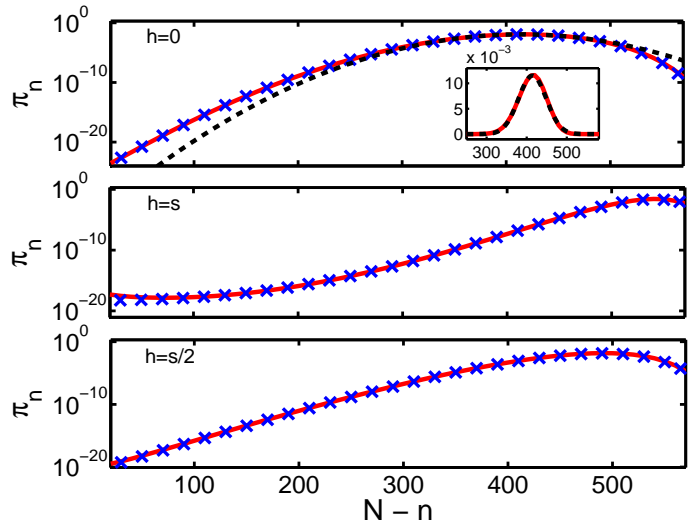}
\caption{Shown is the QSD $\pi_n$ of the number of wild type
in the cases of complete recessivity (top panel), complete dominance (middle panel) and semi-dominance
(bottom panel). The parameters are  ${\cal N}=600, s=0.25$ and $v=0.025$.
The numerical solution of the master equation~(\ref{master})
($\times$ symbols) is compared with the predictions of the WKB approximation (solid curves)
in the ``bulk'' (i.e. away from the states $n={\cal N}$ and $n=0$) given by (\ref{MFT2}), (\ref{MFT1-QSD1}) and (\ref{MFT3-QSD3}).
In all cases an excellent agreement is observed between the WKB and numerics in the regions of applicability of the WKB.
In the top panel, for the sake of comparison we also show a Gaussian approximation of
the QSD (dashed line). Inset of the top panel: the WKB and Gaussian approximations in the {\it vicinity} of the (metastable)
fixed point $n_*$. While it agrees well with the WKB solution  in the close vicinity of $n_*$, the Gaussian approximation exponentially deviates from the WKB and numerical solutions near the tails.}
\label{QSD_comp}
\end{center}
\end{figure}

Using these results, and the fact that $S''(y^*)\simeq s^2/(4v)$, Eqs.~(\ref{qsdfull}) and (\ref{MFTfull}) become in this case
\begin{eqnarray}\label{MFT3-QSD3}
\tau=\frac{2\sqrt{2\pi {\cal N}}}{s\sqrt{v}}e^{{\cal N} \Delta S},\;\;\;\pi(x)=\sqrt{\frac{v}{2\pi {\cal N}}}e^{-{\cal N}[S(x)-S(x^*)]},
\end{eqnarray}
where $S(x)$ is given by Eq.~(\ref{S-semi}) with $x=1-y$ and $\Delta S$ by (\ref{DeltaS-semi}). Similarly as in the completely-recessive case, the WKB results here are valid as long as condition~(\ref{cond}) is satisfied (see Footnote 10).

\section{Discussion of the results and comparison with the diffusion theory}
We now analyze the various results obtained for the QSD and MFT and then
compare the predictions of the WKB treatment for the MFT with those derived by Kimura
using the diffusion theory~(\cite{Kimura1980}).
\subsection{Comparison and analysis of the results for the QSD and MFT}
\begin{figure}
\begin{center}
\includegraphics[width=3.80in,height=2.75in,clip=]{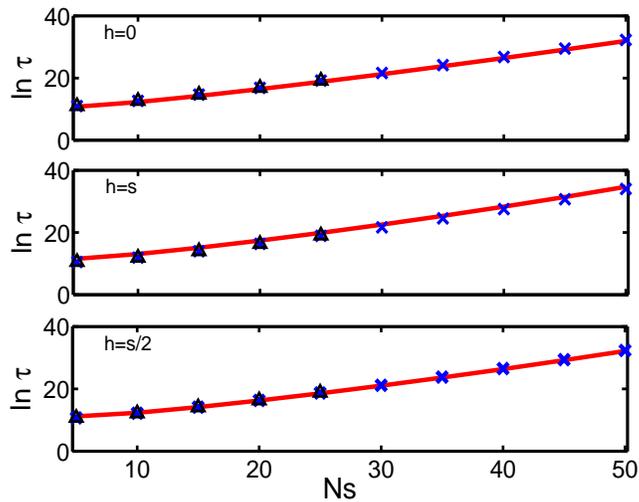}
\caption{Logarithm of the mean fixation time, $\ln{\tau}$, as a function of
${\cal N}s$ for the MM dynamics defined by (\ref{master})-(\ref{trans}) in the cases of complete recessivity
(top panel), complete dominance (middle panel), and semi-dominance (bottom panel).
 The parameters are ${\cal N}=200$, $v=0.005$, while the selection intensity $s$ varies.
The solid lines are the WKB results given by Eqs.~(\ref{MFT2}), (\ref{MFT1-QSD1}), and (\ref{MFT3-QSD3}) for
the top, middle and bottom panels, respectively. The  symbols $\times$ are results of the numerical solution of
 the master equation~(\ref{master}), while the triangles are results obtained from MC
simulations averaged over $100,000$ runs, see Appendix. One can see a very good agreement between the WKB and numerical results. (Here, the MC simulations have been carried out up to $\ln{\tau}\approx 20$ above which they become excessively time consuming.) }
\label{MFT_comp_s}
\end{center}
\end{figure}

\begin{figure}[ht]
\begin{center}
\includegraphics[width=4.0in,height=2.75in,clip=]{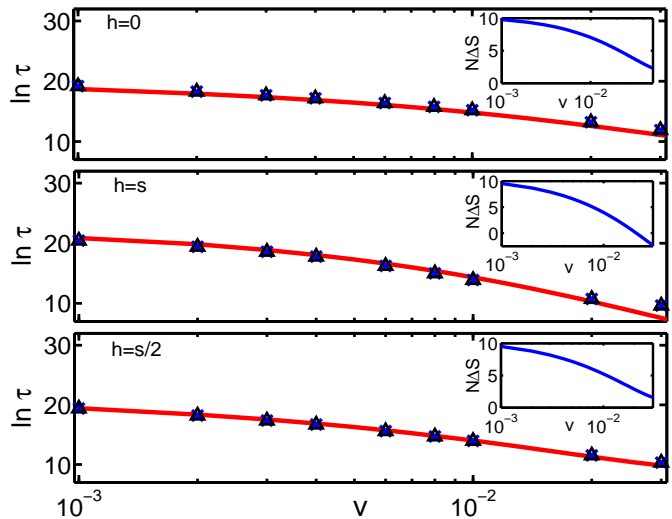}
\caption{Logarithm of the mean fixation time, $\ln{\tau}$, as a function of
mutation rate $v$ for the MM dynamics defined by (\ref{master})-(\ref{trans}) in the cases of complete recessivity
(top panel), complete dominance (middle panel), and semi-dominance (bottom panel).
 The parameters are ${\cal N}=200$ and $s=0.1$. The solid lines are the WKB results given by Eqs.~(\ref{MFT2}), (\ref{MFT1-QSD1}),
 and (\ref{MFT3-QSD3}) for top, middle and bottom panels, respectively, while the $\times$ symbols are the numerical solution of the master equation~(\ref{master}), and the triangles are results of MC simulations
averaged over $100,000$ runs, see Appendix. One can see a very good agreement between the WKB and numerical results. This agreement, however, deteriorates as $v$ is increased such that ${\cal N}\Delta S$ becomes ${\cal O}(1)$, see text. The insets show the dependence of ${\cal N}\Delta S$ on $v$ given by Eq.~(\ref{DeltaS-dom}), (\ref{DeltaS1}), and (\ref{DeltaS-semi}), respectively.}
\label{MFT_comp_v}
\end{center}
\end{figure}

Figures~\ref{QSD_comp} to \ref{MFT_comp_v} summarize the results obtained for the QSD and MFT
in the cases of (i) complete dominance, (ii) complete recessivity, and (iii) semi-dominance of the mutant
allele $\textsf{A}'$. In Fig.~\ref{QSD_comp} we compare the predictions of the WKB theory with the
numerical solution of the master equation~(\ref{master}) and excellent agreement is observed in all three cases.
In particular, in the top panel we
remark that the QSD is bell-shaped and Gaussian only in the close vicinity of the metastable
state $n_*={\cal N}x^*$ (see inset of top panel of Figure~\ref{QSD_comp}), while the tails of the QSD are clearly asymmetric
and  non-Gaussian. Note, that differently from
the QSDs of the top and bottom panels, corresponding to extinction scenario A, the QSD in the middle panel corresponds
to extinction scenario B and displays an \textit{increase} of probability towards the absorbing state $n={\cal N}$~(\cite{AM}).
In this case the WKB approximation breaks down in the vicinity of the repelling fixed  point ${\cal N}y_u^*$,
where another approximation has to be used instead (see Sec.~3).

In Figs.~\ref{MFT_comp_s} and \ref{MFT_comp_v}, we compare the predictions of the WKB results (\ref{MFT2}), (\ref{MFT1-QSD1}),
(\ref{MFT3-QSD3}) for the MFT with those obtained from the numerical solution of the master equation~(\ref{master})
and with MC  simulations averaged over $100,000$ runs. Very good agreement is observed in all cases.
As predicted by our theoretical results, and since $\Delta S$ increases with $s$,  in all three cases, the MFT displays exponential growth
with  ${\cal N}s$ (i.e. the effective
population size multiplied by the selection strength), as found in Fig.~\ref{MFT_comp_s}. Yet, as $v$ is essentially the drift towards the absorbing state,
the MFT decays when the mutation rate $v$ is increased, as reported in Fig.~\ref{MFT_comp_v}. The systematic deviations seen in Fig.~\ref{MFT_comp_v} as $v$ is increased (and approaches the value of $s$) stem from the breakdown of two requirements of the WKB theory, namely ${\cal N}\Delta S\gg 1$ and $v\ll s$ (see insets for the dependence of $\Delta S$ on $v$).

\subsection{Comparison between  the diffusion theory and the WKB treatment}
We now compare our WKB-based predictions for the MFT with
those obtained by Kimura using the diffusion theory [Eqs.~(11)-(13) of Ref.~(\cite{Kimura1980})].
For such a purpose it suffices to focus on the leading contribution to
the MFT and consider its logarithm. We demonstrate that in the subregime $v\ll s\ll {\cal N}^{-1/2}$, the predictions of the diffusion theory coincide (to leading order) with those of our theory, while over a wide range of selection intensity, ${\cal N}^{-1/2}\ll s\ll 1$, the diffusion theory
is plagued by exponentially large errors.

First, it is useful to simplify the result obtained  by Kimura
[see Eq.~(11) of~(\cite{Kimura1980})], which reads
\begin{eqnarray}
 \label{Kimura-MFT}
\tau^{KIM}_{{\rm diff}}(x)={\cal N}\int_{x}^{1}
e^{B(\eta)\eta^{-{\cal N} v}} d\eta \: \int_{0}^{\eta} \frac{e^{-B(\xi)}\xi^{{\cal N} v -1}}{1-\xi} \; d\xi,
\end{eqnarray}
where $B(x)={\cal N} x[s x+2h(1-x)]/2$.
According to the above discussion,  here we have put $4N_e\to 2N_e={\cal N}$ to
map the (``microscopic'') WFM considered  by Kimura to derive
his diffusion theory in (\cite{Kimura1980}) onto the MM dynamics~(\ref{fit1})-(\ref{trans}) discussed in this work.
Our goal here is to evaluate the
asymptotic behavior of (\ref{Kimura-MFT}) in the limit ${\cal N}\gg 1$, $v\ll s\ll 1$, and ${\cal N}s\gg 1$, where our WKB-based
predictions are certainly valid. For this we rewrite Eq.~(\ref{Kimura-MFT}) as
\begin{eqnarray}
 \label{K1}
\tau^{KIM}_{{\rm diff}}(x)={\cal N}\int_{x}^{1} e^{({\cal N}/2) f(\eta)} d\eta \:
\int_{0}^{\eta} \frac{e^{-({\cal N}/2)f(\xi)}}{\xi(1-\xi)} \; d\xi,
\end{eqnarray}
where
\begin{equation}\label{fx}
f(x)=x[s x+2h(1-x)]-2v\ln x.
\end{equation}
As the function in the exponent is rapidly varying, the inner integral can be
evaluated using a saddle-point approximation while
the outer integral is evaluated by using a Taylor expansion.
For the saddle-point calculation, the maximum of the function $-f(\xi)$ is attained at
$\xi^*\simeq\sqrt{v/s}$ for $h=0$, $\xi^*\simeq v/s$ when $h=s$, and $\xi^*\simeq 2v/s$ for $h=s/2$.
We note that $\xi^*$ coincides (to leading order) with the attracting fixed point $x^*$ in
all three cases of interest. Therefore, the inner integral of Eq.~(\ref{K1}) can be approximated
as~\footnote{It is justified to set the boundaries of the inner integral to $\pm \infty$, as the integrand of Eq.~(\ref{K1}) with respect to $\eta$ is regular at $\eta\to 1$, and the peak of the Gaussian is positioned sufficiently far from the boundaries in all three cases $h=0$, $h=s/2$ and $h=s$.} $\int_{0}^{\eta} [\xi(1-\xi)]^{-1}e^{-({\cal N}/2) f(\xi)} d\xi\simeq [\xi^*(1-\xi^*)]^{-1}\int_{-\infty}^{\infty} e^{-({\cal N}/2) f(\xi^*)-({\cal N}/4)f''(\xi^*)
(\xi-\xi^*)^2} d\xi\sim e^{-({\cal N}/2)f(\xi^*)}$.
As this integral turns out to be independent of $\eta$,
and $f(\xi)$ is an increasing function on $\xi^*\leq\xi\leq 1$,
the main contribution to the remaining outer integral of (\ref{K1}) arises from $\eta=1$.
Thus, Taylor-expanding the integrand about $\eta=1$,
to leading order the remaining integral becomes $\int_{x}^{1} e^{({\cal N}/2) f(\eta)} d\eta\simeq \int_{x}^{1} e^{({\cal N}/2) f(1)-({\cal N}/2)
(1-\eta)f'(1)} d\eta\sim e^{({\cal N}/2) f(1)}$.
Therefore, using (\ref{fx}), substituting the above integrals into (\ref{K1}), and neglecting logarithmic corrections, Kimura's result (\ref{Kimura-MFT})
becomes:
\begin{equation}\label{taukim}
\ln \tau^{KIM}_{{\rm diff}}(x) \simeq ({\cal N}/2)[f(1)-f(x^*)].
\end{equation}

We now consider the predictions of the WKB theory for
the logarithm of the MFT in the limit $v \ll s \ll 1$ (with $s\gg {\cal N}^{-1}$ to ensure the validity of
the WKB treatment). It follows from Eq.~(\ref{mainMFT}) that
\begin{equation}\label{tauwkb}
\ln\tau_{WKB}\simeq {\cal N} \Delta S = {\cal N} [S(1)-S(x^*)]
\end{equation}
for all three cases (i)-(iii) considered here. For the sake of comparison with the predictions of the diffusion
theory, with (\ref{Ssingle}) and (\ref{tau-x}), we notice that  to linear order in $s$ and $h$ one has
\begin{equation}\label{actsmalls}
S(x)=\int^x \ln \left(\frac{T^-(\xi)}{T^+(\xi)}\right)d\xi\simeq x\left[\frac{sx}{2}+h(1-x)\right]- v\ln x+{\cal O}(s^2, h^2, s\,h).
\end{equation}
From Eqs.~(\ref{actsmalls}) and (\ref{fx}), it is clear that the action   $S(x)$ is related to  $f(x)$
by  $S(x)=f(x)/2$.
It follows from this discussion that in the limit $ v\ll s\ll {\cal N}^{-1/2}$ (with ${\cal N}s\gg 1$)
the leading-order logarithm of the WKB result (\ref{tauwkb}) exactly coincides with the result~(\ref{taukim}) obtained by Kimura
using the diffusion theory~(\cite{Kimura1980}).

\begin{figure}
\begin{center}
\includegraphics[width=4.5in,height=2.75in,clip=]{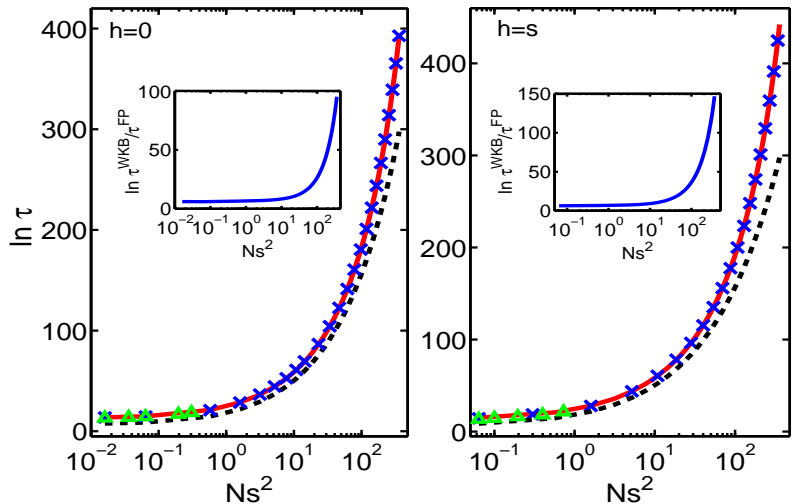}
\caption{Logarithm of the MFT, $\ln{\tau}$,
as a function of ${\cal N}s^2$
in the case of
recessive mutants (left panel), and dominant mutants (right panel). The parameters are
${\cal N}=1000$, $v=0.001$, and
$s=h>0$ varies. The solid curves correspond to the WKB result (\ref{mainMFT}) $\ln{\tau}\simeq {\cal N} \Delta S$ with
 (\ref{DeltaS-dom}). The $\times$ symbols are the results obtained from the numerical solution of the master equation~(\ref{master}), the triangles are MC simulations averaged over $100,000$ runs (see Appendix), while
the dashed curves correspond to the result (\ref{Kimura-MFT}) obtained from the diffusion theory.
The WKB and numerical solutions compare excellently over the entire range of parameters, while the results of diffusion
theory are found to fairly agree with the others for ${\cal N}s^2\lesssim {\cal O}(1)$. However, for ${\cal N}s^2\gg 1$, the dashed curves deviate systematically from the others; in this regime, the diffusion approximation exponentially deviates from the WKB approximation and numerics. Insets: ratio of $\ln{\tau_{WKB}}$ and $\ln{\tau_{FP}}$ illustrating that the
deviations between $\tau_{WKB}$ and $\tau_{FP}$ increase exponentially when ${\cal N}s^2\gg 1$.}
\label{CompDom}
\end{center}
\end{figure}

However, since the WKB is applicable for any $s\ll 1$, one can calculate the next-order corrections to the exponent
of the MFT~(\ref{tauwkb}) and this indeed leads to the following accumulated action:
\begin{eqnarray}
\label{Sbar}
 \Delta S = \left\{
    \begin{array}{ll}
S(1)-S(x^*)\simeq\frac{s}{2}-\frac{v}{2}\left[1-\ln{(v/s)}\right]+\frac{s^2}{3} & \mbox{for} \; h=0\\
S(x_u^*)-S(x^*)\simeq \frac{s}{2}-v\left[1-\ln{(v/s)}\right]+\frac{s^2}{6} & \mbox{for} \; h=s \\
S(1)-S(x^*)\simeq\frac{s}{2}-v\left[1-\ln{(2v/s)}\right]+\frac{s^2}{4} & \mbox{for} \; h=s/2,
    \end{array}
\right.
\end{eqnarray}
where we have used Eqs.~(\ref{S2}), (\ref{S1}), (\ref{S-semi}), and (\ref{actsmalls}) and the expressions of $x^*=1-y^*$ and $x_u^*=1-y_u^*$ determined in
Section 4.
From Eqs.~(\ref{taukim}) and (\ref{Sbar}), one can see that already the next-order corrections ${\cal O}(s^2,h^2,sh)$ are not captured by the diffusion approximation, and thus, the difference between the predictions of the diffusion and WKB predictions
is  given in the leading order by $\ln{\tau_{WKB}}-\ln{\tau_{Kim}}\sim {\cal N}s^2$.
It is thus clear that when ${\cal N}s^2\gg 1$ (or $s\gg {\cal N}^{-1/2}$) there are {\it exponentially large deviations
 between the predictions  of the diffusion theory and those of the WKB approach}. On the other hand, as illustrated by
Figs.~\ref{MFT_comp_s}-\ref{CompDom}, the results of our WKB treatment
are in excellent agreement with numerics over the entire region of parameters.
Therefore, our theory substantiates and quantitatively supports Nei's heuristic argument~(\cite{Nei}).

These findings are illustrated in Fig.~\ref{CompDom} for the case of completely dominant and completely recessive mutants. [Similar results (not shown here) are also obtained for the case of semi-dominance ($h=s/2>0$)]. In Fig.~\ref{CompDom}, the WKB-based results $\ln{\tau_{WKB}}$
are compared with the logarithm of the expression  (\ref{Kimura-MFT}) obtained from the diffusion theory,  as well as with
the numerical solution of the master equation~(\ref{master}) and MC simulations.
From this comparison, it appears that in the regime where $v\ll s \ll {\cal N}^{-1/2}$ the diffusion theory is in fair agreement with the predictions of the WKB theory and numerical simulations. However, for $s\gg {\cal N}^{-1/2}$, the predictions of the diffusion theory are plagued by exponentially large errors.

\section{Discussion and Conclusion}
In this work we have considered a model
 motivated by the problem of fixation in systems experiencing
amorphic or hypermorphic mutations, and studied
the dynamics of a
panmictic diallelic population of diploid individuals at a single locus subject to (small) mutation pressure
from a deleterious allele.
Such a model is characterized by {\it metastability} and its long-time dynamics is therefore governed by large fluctuations.
In this case, the system fluctuates around the metastable state (forming a quasi-stationary distribution about it) until
it is driven into the absorbing state where the population is composed of only the mutant allele.
Since the rare large fluctuations that trigger fixation in this system are ill-described by
the diffusion approximation that was previously used to study this problem~(\cite{Kimura1980}), we have here adopted a different approach based on the WKB theory. This theory has allowed us to
accurately determine the system's mean fixation time (MFT) and quasi-stationary distribution (QSD) for arbitrary
(finite) selection strength and weak mutation rate.

Our treatment is based on a stochastic formulation of the dynamics
in terms of  a birth-death process (Markov chain),
where the transition rates are given by a frequency-dependent version of the Moran model~(\cite{Moran58,Moran62}),
which is closely related to the Wright-Fisher model~(\cite{Wright31,Fisher}) [see Section 2.2].
The master equation  associated with the birth-death process has been treated using the WKB approximation, which is a power-series expansion
in the (effective) population size based on an exponential ansatz (\cite{Landau}).
Using this approach we have investigated the fixation phenomenon when the deleterious mutant allele is
(i) completely dominant,
(ii) completely recessive, (iii) semi-dominant, and our main findings are the following:
\begin{itemize}
 \item We have analytically calculated the MFT  for scenarios (i)-(iii),
including the subleading-order corrections to the MFT that were found to scale as some power of ${\cal N}$.
Our predictions are found to be in excellent agreement
with numerical solutions of the master equation~(\ref{master}), and stochastic Monte Carlo simulations (see Appendix). In all cases we have found that the MFT's exponent grows monotonically with ${\cal N}s$, where ${\cal N}\gg 1$ is
 the effective population size (up to a constant factor) and $s$ is the mutant allele's selection intensity. We have also found that the MFT
decreases when the mutation rate increases [see Eq.~(\ref{Sbar})].
\item We have analytically calculated the QSD up to subleading order for the cases (i)-(iii), and verified our results by comparing them to numerical simulations. In all cases the QSD, centered around the metastable state, is found to be markedly non-Gaussian
when $s$ is nonvanishingly small. We have also found that the shape of the QSD's in cases (ii)-(iii) differs from that of case (i), as cases (ii) and (iii) correspond to a different fixation scenario than case (i)~(\cite{AM}).
\item Our (leading order) predictions for the MFT, based on the WKB theory, have been compared with Kimura's
predictions obtained from the diffusion approximation~(\cite{Kimura}). This comparison indicates that the diffusion approximation is valid for ${\cal N}s^2\ll 1$. Yet, when nonlinear contributions of the selection strength are
no longer negligible, the predictions of the diffusion approximation are found to be exponentially flawed and their inaccuracy grows with the (effective) population size, as illustrated by Fig.~\ref{CompDom}. This demonstrates that  ${\cal N}s^2\ll 1$
is the {\it weak selection limit} where the diffusion theory is an adequate approximation, which substantiates
a recent heuristic argument  (\cite{Nei}).
\end{itemize}
Finally, these results shed further light on the interplay between genetic drift and selection. They also illustrate that in systems exhibiting
metastability and governed by rare large fluctuations, the diffusion theory is well-suited only in a narrow range of parameters (weak selection  limit) where the dynamics is almost neutral.
\section*{Acknowledgments}
M.~A. would like to acknowledge support from the Rothschild and Fulbright Foundations.

\begin{appendix}

\section*{Appendix:~Description of the stochastic simulations }

In this Appendix we briefly outline the stochastic Monte Carlo (MC) method, using an algorithm due to Gillespie~(\cite{Gillespie77}), which is often used to simulate stochastic
birth-death processes.
A MC simulation yields a single realization (or run) of the stochastic dynamics of the population.
Clearly, in order to obtain the PDF of population sizes, one has to calculate the corresponding histogram by the use of binning.

For the sake of completeness, the MC-Gillespie algorithm is here briefly illustrated in the   case of
a well-mixed (randomly-mating) population that is of direct relevance
for our purposes. As before, we consider birth and death rates respectively given by
 $T^+_n$ and  $T^-_n$.
Starting with $n_0$ mutant alleles, the following loop has
to be repeated until the number of mutant alleles reaches the absorbing state:
\begin{itemize}
\item  Calculate the probabilities for birth and death in the next time step.
The probability that birth occurs is $P_{b}=T^+_n/[T^+_n+T^-_n]$, while the probability of death is
$P_{d}=1-P_{b}$, where $n$ is the current number of mutants $\textsf{A}'$.
\item  Update the physical time of the next step. This time is drawn from an exponential distribution
with mean that equals the inverse of the sum of the birth and death reaction rates. That is, the probability that
a time $\Delta t$ has elapsed since the last step satisfies: $P(\Delta t)=(1/{\cal T})e^{-\Delta t/{\cal T}}$,
where ${\cal T}=[T^+_n+T^-_n]^{-1}$.
\item  Update number of mutants. Choose a random number $c$ between $0$ and $1$. If $c<P_{b}$, birth occurs and $n$ increases by $1$;
otherwise death occurs and $n$ decreases by $1$.
\item  If the updated number of mutants $n$ reaches one of the absorbing states, exit the loop. Otherwise go back to the first step.
\end{itemize}
This algorithm generates a single erratic trajectory that mirrors
the stochastic dynamics of the population.
The time in these simulations is the real physical time, and therefore,
averaging over the fixation time of all these trajectories yields the MFT of the process.

It is worth noticing that the both the above simulation algorithm and the pseudo-sampling method used in ~(\cite{Kimura1980})
 keep track of the current number of mutants $n(t)$ by drawing a random number at each time
step and by sequentially updating $n(t)$.
However, while in our simulations
$n(t)$ is updated by $\pm 1$ at each time step according to the current probability to undergo birth or death,
Kimura's algorithm updates $n(t)$ by adding to it a random number with mean zero and variance that coincides with that of
 the underlying diffusion process. Hence,
while our simulations
exactly mirror the (full) underlying evolutionary stochastic process (whose PDF obeys the master equation (\ref{master})),
 Kimura's pseudo-sampling method replicates the predictions of the  diffusion approximation (\ref{bKE}) and is
therefore expected to be accurate only in the limit of weak selection intensity.

\end{appendix}


\begin{thebibliography}{70}
\expandafter\ifx\csname natexlab\endcsname\relax\def\natexlab#1{#1}\fi
\expandafter\ifx\csname url\endcsname\relax
  \def\url#1{\texttt{#1}}\fi
\expandafter\ifx\csname urlprefix\endcsname\relax\def\urlprefix{URL }\fi
\providecommand{\selectlanguage}[1]{\relax}

\bibitem[{Assaf and Meerson(2006a)}]{AM0}
{Assaf, M., Meerson, B.}, 2006.
\newblock Spectral formulation and WKB approximation for rare-event statistics in reaction systems.
\newblock Phys. Rev. E {74}, 041115.

\bibitem[{Assaf and Meerson(2006b)}]{AM1}
{Assaf, M., Meerson, B.}, 2006.
\newblock  Spectral theory of metastability and extinction in birth-death systems.
\newblock Phys. Rev. Lett. {97}, 200602.

\bibitem[{Assaf and Meerson(2007)}]{AM2}
{Assaf, M., Meerson, B.}, 2007.
\newblock  Spectral theory of metastability and extinction in a branching-annihilation reaction.
\newblock Phys. Rev. E {75}, 031122.

\bibitem[{Assaf and Meerson(2010)}]{AM}
{Assaf, M., Meerson, B.}, 2010.
\newblock Extinction of metastable stochastic populations.
\newblock Phys. Rev. E {81}, 021116.

\bibitem[{Assaf and Mobilia(2010)}]{MAMM}
{Assaf, M., Mobilia, M.}, 2010.
\newblock Large Fluctuations and Fixation in Evolutionary Games.
\newblock J. Stat. Mech. {\bf P09009}

\bibitem[{Blythe and McKane(2007)}]{Blythe}
{Blythe, R.~A., McKane, A.~J.}, 2007.
\newblock Stochastic models of evolution in genetics, ecology and linguistics.
\newblock  J. Stat. Mech.  {P07018}.

\bibitem[{Crow and Kimura(1970)}]{CrowKimura}
{Crow, J.~F., Kimura, M.}, 1970.
\newblock An Introduction to Population Genetics Theory.
\newblock Harper and Row, New York.

\bibitem[{Dykman et al.(1994)}]{Dykman}
{Dykman, M.~I., Mori, E., Ross, J., Hunt, P.~M.}, 1994.
\newblock Large fluctuations and optimal paths in chemical kinetics.
\newblock J. Chem. Phys. {100}, 5735.

\bibitem[{Durrett and Schweinsberg(2004)}]{Durrett}
{Durrett, R. and Schweinsberg, J.}, 2004.
\newblock Approximating selective sweeps.
\newblock Theor. Popul. Biol. {66}, 129.

\bibitem[{Eriksson et al.(2008)}]{Eriksson}
{Eriksson, A., Fernstr\"om, P., Mehlig, B., and Sagitov, S.}, 2008.
\newblock An accurate model for genetic hitchhiking.
\newblock Ann. Appl. Probab. {16}, 685.

\bibitem[{Escudero and Kamenev(2009)}]{EsK}
{Escudero, C, Kamenev, A}, 2009.
\newblock Switching rates of multistep reactions
\newblock Phys. Rev. E {79}, 041149.

\bibitem[{Etheridge et al.(2006)}]{Etheridge}
{Etheridge, A., Pfaffelhuber, P., and Wakolbinger, A.}, 2006.
\newblock An approximate sampling formula under genetic hitchhiking.
\newblock Ann. Appl. Probab. {16}, 685.

\bibitem[{Ewens(2000)}]{Ewens}
{Ewens, W.~J.}, 2000.
\newblock Mathematical Population Genetics. I. Theoretical
Introduction.
\newblock Springer, New York, 2nd edition.

\bibitem[{Feller(1968)}]{Feller}
{Feller, W.}, 1968.
\newblock An Introduction to Probability Theory and its Application, vol.~1.
\newblock Wiley, London, third ed.

\bibitem[{Fisher(1922)}]{Fisher}
{Fisher, R.~A.}, 1922.
\newblock On the dominance ratio.
\newblock Proc. R. Soc. Edinb. {42}, 321.

\bibitem[{Fisher(1930)}]{PopGen}
{Fisher, R.~A.}, 1930.
\newblock The Genetical Theory of Natural Selection.
\newblock Clarendon Press, Oxford, U.K.

\bibitem[{Gardiner(2002)}]{Gardiner}
{Gardiner, C.~W.}, 2002.
\newblock Handbook of Stochastic Methods.
\newblock Springer Verlag, Berlin, 2nd ed.

\bibitem[{Gillespie(2004)}]{Gillespie04}
{Gillespie, J.~H.}, 2004
\newblock Population genetics: a concise guide.
\newblock Johns Hopkins University Press, Baltimore.

\bibitem[{Gillespie(1977)}]{Gillespie77}
{Gillespie, D.~T.}, 1977
\newblock Exact Stochastic Simulation of Coupled Chemical Reactions.
\newblock J. Phys. Chem. {81}, 2340.

\bibitem[{Jukes and King(1975)}]{Jukes1975}
{Jukes, T.~H., King, J.~L.}, 1975.
\newblock Evolutionary loss of ascorbic acid synthesizing ability.
\newblock J. Hum. Evol. {4}, 85.


\bibitem[{Kimura(1980)}]{Kimura1980}
 {Kimura, M.}, 1980.
\newblock Time until Fixation of a Mutant Allele in a Finite Population under Continued Mutation Pressure.
\newblock Proc. Natl. Acad. Sci. U.S.A. {77}, 522.

\bibitem[{Kimura and Ohta(1971)}]{KimuraOhta}
{Kimura, M., Ohta, T.}, 1971.
\newblock Theoretical Aspects of Population Genetics.
\newblock Princeton University Press, Princeton.

\bibitem[{Kimura(1983)}]{Kimura}
{Kimura, M., Ohta, T.}, 1983.
\newblock The Neutral Theory of Molecular Evolution.
\newblock Cambridge University Press, Cambridge, U.K.

\bibitem[{Korolev et al.(2010)}]{Korolev}
{Korolev K. S., Avlund, M., Hallatschek, O., Nelson, D.~R.}, 2010
\newblock  Genetic demixing and evolution in linear stepping stone models.
\newblock  Rev. Mod. Phys. {82}, 1691.

\bibitem[{Kubo et al.(1973)}]{kubo}
{Kubo, R., Matsuo, K., Kitahara, K.}, 1973.
\newblock J.~Stat.~Phys. {9}, 51.
\newblock Fluctuation and relaxation of macrovariables

\bibitem[{Landau and Lifshitz(1977)}]{Landau}
{Landau, L.~D., Lifshitz, E.~M.}, 1977.
\newblock Quantum Mechanics:Non-Relativistic Theory.
\newblock Pergamon, London.

\bibitem[{Li and Nei(1977)}]{Li}
{Li, W.-H., Nei, M.}, 1977.
\newblock Persistence of common alleles in two related populations or species
\newblock Genetics {86}, 901.

\bibitem[{Mobilia and Assaf(2010)}]{MA}
{Mobilia, M., Assaf, M.}, 2010.
\newblock Fixation in Evolutionary Games under Non-Vanishing Selection.
\newblock EPL {91}, (10002).


\bibitem[{Moran(1958)}]{Moran58}
{Moran, P. A.~P.}, 1958.
\newblock Random processes in genetics.
\newblock Proc. Camb. Phil. Soc. {54}, 60.

\bibitem[{Moran(1962)}]{Moran62}
{Moran, P.~A.~P.}, 1962.
\newblock The Statistical Processes of Evolutionary Theory.
\newblock Clarendon Press, Oxford, U.K.

\bibitem[{Muller(1939)}]{Muller1939}
{Muller, H. J.}, 1939.
\newblock  Reversibility in evolution considered from the standpoint of genetics.
\newblock Biol. Rev., Cambridge {14}, 261.

\bibitem[{Nei(2005)}]{Nei}
{Nei, M}, 2005.
\newblock Selectionism and Neutralism in Molecular Evolution.
\newblock Mol.~Biol.Evol. {22}, 2318.



\bibitem[{Risken (1989)}]{Risken}
{Risken, H.}, 1989.
\newblock The Fokker-Planck Equation.
\newblock Springer, second ed.

\bibitem[{Sella and Hirsh(2005)}]{Sella}
{Sella, G. and Hirsch, A.~E.}, 2005.
\newblock The application of statiscal physics to evolutionary biology.
\newblock Proc. Natl. Acad. Sci. U.S.A. {102}, 954




\bibitem[{Slatkin and Rannala(2000)}]{Slatkin}
{Slatkin, M., Rannala, B.}, 2000.
\newblock Estimating allele age.
\newblock Genomics Hum. Genet. {1}, 225.
%

\bibitem[{van Kampen(1992)}]{VanKampen}
{van Kampen, N.~G.}, 1992.
\newblock Stochastic processes in physics and chemistry.
\newblock North Holland, Amsterdam, second edition.

\bibitem[{Wang and Rannala(2004)}]{Wang}
{Wang, Y., Rannala, B.}, 2004.
\newblock A novel solution for the time-dependent probability of gene fixation or loss under natural selection.
\newblock Genetics {168}, 1081.

\bibitem[{Whitlock(2003)}]{Whitlock}
{Whitlock, M.~C.}, 2003.
\newblock Fixation Probability and Time in Subdivided Populations.
\newblock Genetics {164}, 767.



\bibitem[{Wright(1977)}]{Wright1977}
{Wright, S}, 1977.
\newblock  Evolution and the Genetics of Populations.
\newblock Univ. of Chicago press, Chicago, 1977, Vol.3.

\bibitem[{Wright(1931)}]{Wright31}
{Wright, S.}, 1931.
\newblock Evolution in Mendelian Populations.
\newblock Genetics {16}, 97.

\bibitem[{Zeng et al.(2007a)}]{Zeng1}
{Zeng, K., Mano, S., Shi, S., Wu, C.-I.}, 2007(a).
\newblock Comparisons of Site- and Haplotype-Frequency Methods for Detecting Positive Selection.
\newblock Mol. Biol. Evol. {24}, 1562.

\bibitem[{Zeng et al.(2007b)}]{Zeng2}
{Zeng, K., Shi, S., Wu, C.-I.}, 2007(b).
\newblock Compound Tests for the Detection of Hitchhiking Under Positive Selection.
\newblock Mol. Biol. Evol. {24}, 1898.






\end{thebibliography}
\end{document}